\documentclass{article}
\usepackage[utf8]{inputenc} % Portugueses
\usepackage{amsmath} % math package
\usepackage{amssymb} % math package
\usepackage{booktabs}
\usepackage{fullpage} % minor margins
\usepackage{graphicx} % include figure
\usepackage{verbatim}
\usepackage{url}
\usepackage{caption}
\usepackage{subcaption}
\usepackage{booktabs}
\usepackage{floatrow}
\usepackage{color}

\title{Spatial distribution of graffiti: A complex network approach}

\author{Eric K. Tokuda$^{1}$ \and Henrique F. de Arruda$^{1}$ \and Cesar H. Comin$^{2}$ \and Roberto M. Cesar-Jr.$^{1}$ \and Claudio T. Silva$^{3}$ \and Luciano da F. Costa$^{1}$}

\date{%
    $^1$ University of São Paulo, Brazil \\
    $^2$ Federal University of São Carlos, Brazil \\
    $^3$ New York University, USA
}

\begin{document}
\maketitle

\abstract{Despite the great differences among cities, they face similar challenges regarding social inequality, politics and criminality. Urban art express these feelings from the citizen point-of-view. In particular, the drawing and painting of public surfaces may carry rich information about the time and region it was made.  Existing studies have explored the spatial distribution of graffiti, but most of them considered graffiti as a whole, with no separation among the types. Also, the analyses rarely take into account the city topology. In this work, we propose to categorize the graffiti into three types: simple scribbles, complex scribbles, canvases. We analyze the spatial distribution and identify the spatial bias of each type. To further analyze the spatial distribution of the types, we apply concepts from complex networks. First, regions (communities) defined by the connectivity profiles of the city network are obtained and the prevalence of each type of graffiti over these regions are analyzed. Next, a measure based on the dynamics of the network (accessibility) is calculated and compared to the distribution of the graffiti types. A case study is performed in the analysis of three different categories of graffiti in the city of São Paulo, Brazil. The results showed that the categories present characteristic spatial distributions. The ratio of each type per community of the network, though, does not pose significant deviations. Finally, a small positive correlation was observed between the locations of each graffiti type and the accessibility.}

\section{Introduction}
	Modern cities represent the perfect stage for different forms of urban expression. Paintings on public walls are commonly known as \emph{graffiti}, being part of the landscape of most large cities. These \emph{urban canvases} may be composed not just by drawings, but also by combinations of symbols and writings. There is a wide range of ways in which these graffiti can be made, regarding the material, the size, the colour, and the respective content. These visual signature may be associated with the socio-economical context of the location~\cite{ferrell1993crimes}. Despite their great diversity, 
	for simplicity's sake here we call all these forms graffiti.

	It can be observed that graffiti is primarily created in urban regions. From a mathematical point of view, cities can be seen as an irregular, complex graph that evolves in time~\cite{barthelemy2011spatial}.  As such, they can be approached by using complex network concepts and methods. A diverse range of properties can be obtained from the network structure, including its average degree, clustering coefficient and centrality measures. Interestingly, these topological features can be related to
	several dynamical processes taking place in the network, such as
	mobility.
	
    Previous works have analyzed the distribution of graffiti and its correlation with socio-economical indices~\cite{haworth2013spatio,tokuda2019quantifying,megler2014spatial,alonso1998urban}. These works considered just particular cases of graffiti, and more systematic studies considering the spatial distribution of different categories of graffiti are still necessary.  Another subject deserving further attention consists in trying to better understand graffiti in the light of the city network topology and dynamics. In this work, we aim at approaching these problems in more systematic way.

	Street-level images are initially collected from the location of interest. Categories are then defined based on a visual exploration of the data. The collected images are classified in terms of the observed categories of graffiti. The distribution of the location of each type is calculated and its deviation to the overall distribution is estimated. To investigate these differences, we extract the street network and analyze the respective structural and dynamics properties. From the topology, we obtain the communities and analyze the difference in the participation of each type of graffiti across the communities. We additionally perform random walks in the network in order to estimate the \emph{accessibility} of the nodes. The results suggest differences among the distributions of each type of graffiti.  Finally, the results of the dynamics show that there are moderate positive correlations among the density of each type and the accessibility.

	The paper is organized as follows: the method and the underlying theory are initially presented; the experiments are described; followed by the respective conclusions.

\section{Materials and Methods}
Instead of manually inspecting the regions in the city, the proposed approach consider street-level images collected from Google Street View~\cite{googlemaps}. %
These images are obtained from sensor-equipped cars as they move along
the streets, with similar weather and lighting of the scene within a given region. The images are accompanied by metadata, which includes the geo-location of the scene. In the presented method, a geographical region of interest is initially defined (e.g.~corresponding to a specific city). Given that a very large number of images is usually found in~\cite{googlemaps}, it often becomes a challenge to process all of them.  In this work, we randomly sample, with uniform distribution, a percentage of these images.  As the images available in~\cite{googlemaps} do not contain metadata regarding the presence of graffiti, it is necessary to perform some pre-processing in order to separate the images that are to be considered.

When the the overall distribution of graffiti is not uniform, presenting variations of density at distinct positions,  two main hypotheses may be considered to explain such heterogeneity distributions: (i) though originally distributed in uniform manner,  the graffiti was sampled in some heterogeneous manner; and (ii) the graffiti is, in fact, not uniformly distributed.  A third possibility would correspond to a combination of these two hypotheses, namely a heterogeneous graffiti distribution being sampled in a non-uniform manner.  For simplicity's sake,  hypothesis (i) was adopted in the present work. Such approach is associated to a choice of a minimum density. Therefore, it becomes necessary to consider some indication of relevance that allows the less densely sampled portions of the data not to be taken into account.  This criterion is described in more detail in \emph{Experiments}.

\begin{figure}[ht!]
    \begin{subfigure}[b]{0.3\textwidth}
	    \centering
	    \includegraphics[width=\textwidth]{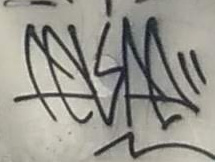}
	\caption{Type A}
    \end{subfigure}
    \begin{subfigure}[b]{0.3\textwidth}
	    \centering
	\includegraphics[width=\textwidth]{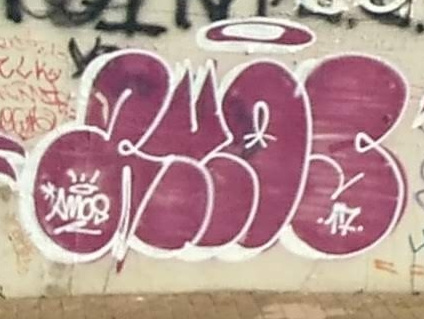}
	\caption{Type B}
    \end{subfigure}
    \begin{subfigure}[b]{0.3\textwidth}
	    \centering
	\includegraphics[width=\textwidth]{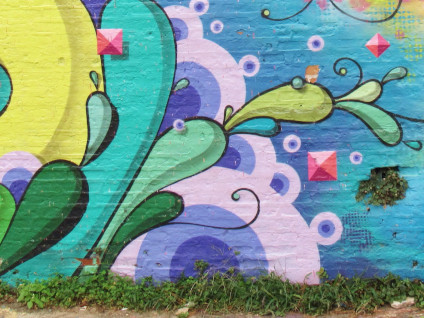}
	\caption{Type C}
    \end{subfigure}
    \caption{Categories of graffiti considered in this work
    (from a personal collection of pictures).}
    \label{fig:graffititypes}
\end{figure}

Differently from previous related works, the graffiti present in the image is categorized according to their visual form: simple scribbles (type A), painted writings (type B), and canvases or textures (type C). This division is made after a preliminary visual exploration of the graffiti occurrences, taking into account the expected time to make them, while taking care to achieving well-separated categories. Please refer to Figure~\ref{fig:graffititypes} for an illustration of each of these three types. The collected sample is manually labeled whether containing or not each of the three types described, resulting in $\sum_{m=0}^{3} {3 \choose m} = 8$ possible combinations of labels for an image.

\subsection{Definition of types and annotation}

To compare the spatial distribution of each of the above types of graffiti, a non-parametric density estimation with smoothing kernel can be calculated on the location of the occurrences. A kernel density estimation~\cite{davis2011remarks} with gaussian kernel is therefore computed for each of the occurrence types. To compare how the distributions \emph{differ} one another, the Kullback-Leibler (KL) divergence~\cite{kullback1997information} is calculated. The KL divergence is a non-symmetric measure that originated from information theory, to estimate the information lost between the expected (Q) and the observed signal (P). It is calculated as $ D_\text{KL}(p(x) \parallel q(x)) = \int_{-\infty}^\infty p(x) \log\left(\frac{p(x)}{q(x)}\right)\, dx $. According to the equation, lower values corresponding to \emph{similar} distributions.

To analyze the influence of the streets network topology over the distribution of each type of graffiti, topological network information from the same region is obtained. OpenStreetMap~\cite{openstreetmap}, a collaborative project, was considered for that purpose.

\subsection{Network structure}
Real networks~\cite{boccaletti2006complex} are characterized by presenting heterogeneous connectivity patterns. One classical problem when handling graphs is \emph{community detection}, which aims at identifying modules or subgraphs having vertices that are more intensely interconnected internally to each module than across modules. Many works have tackled this problem~\cite{fortunato2010community}, including the more traditional partition-based, hierarchical-based, and spectral-based methods. Instead of handling expensive computations on the adjacency matrices, Rosvall et al.~\cite{rosvall2008maps} proposed a community detection methodology based on random walks and information diffusion. The method minimizes the information required to describe the process of information diffusion across the graph. For instance, in a two-level description, communities in the graph may be uniquely identified, while the name of the vertices may be recycled across these structures. This idea is analogous to repeating street names across different cities. Huffman coding is used for the vertex names and the method aims at finding the partitioning of the graph which minimizes the description length of an infinite walk. In the case of no well-defined clusters, transitions between the potential clusters will be frequent and there will be little benefit in using a two-level description. The optimization of the possible partitioning schemes is computed by using greedy search and simulated annealing. The described approach is applied to the identification of clusters in the city network.

\subsection{Network dynamics}
The structure of a network plays an important role in the dynamics of the processes that can take place on it (e.g.~\cite{boccaletti2006complex}).  \emph{Accessibility}~\cite{travenccolo2008accessibility} is an interesting concept in this regard, as it provides information about both the structure and dynamics of complex networks. This measurement quantifies the influence of a given node on a set of vertices in its neighborhood, with respect to a specific dynamics (e.g.~neuronal dynamics, epidemics, random walks, etc.). The random walk represents one of the central paradigms in dynamical systems, partly due to its simple formulation. The dynamics on random walks have been extensively studied (e.g.~\cite{newman2006structure}) in the literature. 

\begin{figure}[ht]
	\centering
	\includegraphics[width=0.6\textwidth]{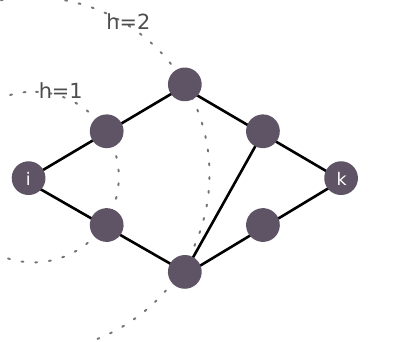}
	\caption{Reachable vertices in the accessibility measurement. Concentric vertices represent nodes reachable in the same number of steps, h=1,2, departing from node i. This measure achieves a higher value for vertex i than for vertex j, assuming h=2.}
	\label{fig:accessibdiagram}
\end{figure}

In this  work, a variation of this classical process, in which the walker is not allowed to repeat any vertex or edge is considered because it is not expected that the graffiti makers keep going back and forth in the same streets. The agent starts from a source node and is allowed to move in a walk with no repeating vertices or edges, which results in a path with length $h$ (please refer to Figure~\ref{fig:accessibdiagram}). The transition probabilities ($p^h_{i,j}$) between vertices can be obtained by considering the number of times the vertex $j$ was reached from vertex $i$, $c^h_{i,j}$ scaled by the number of realizations (number of paths). The outward accessibility of each vertex can then be calculated with Equation~\ref{eq:accessib}. 
\begin{equation}
    A^{out}_{h}(i) = exp \left[- \sum^{n}_{j=1} p^h_{i,j} log (p^h_{i,j}) \right].
	\label{eq:accessib}
\end{equation}

This measure reaches maximum value when it can reach the vertices in level $h$ with the same probability. Consider the network depicted in Figure~\ref{fig:accessibdiagram}. In this case, the accessibility of the vertex i for $h=2$ has maximum value, $A_{h=2}^{out}(i)=exp \left[- (\frac{1}{2} log (\frac{1}{2})+ \frac{1}{2} log (\frac{1}{2})\right]=2$, because it can reach the vertices in level 2 with the same probability. On the other hand, this same measure achieves a lower value for vertex $k$ due to the different probabilities of reaching the vertices 2 levels apart from $k$.

\section{Experiments}
\label{sec:experiments}
We considered the city of São Paulo, Brazil as a case study. We initially obtained from~\cite{openstreetmap} the street network, which is directed and composed of 112,014 nodes and 287,592 edges.

\begin{figure}
    \centering
	\includegraphics[width=.6\textwidth]{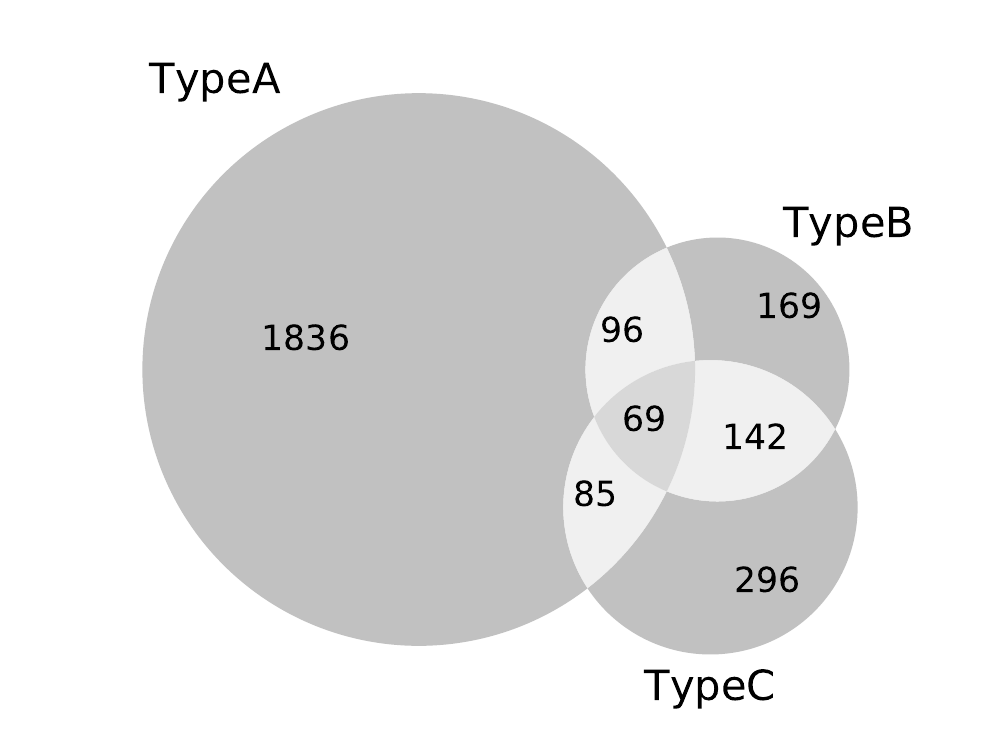}
    \caption{Images with occurrences of graffiti per type. Different types may be present in a single image.}
    \label{fig:coocurrence}
\end{figure}

A total of 16,000 images was considered in this work, which consisted of a uniformly random sample of roughly 275,000 images from~\cite{googlemaps} the city of São Paulo. The images were initially obtained and manually annotated in terms of the three types of graffiti or their absence. A total of 3,154 images were identified as containing at least one type of graffiti. The numbers of occurrences of graffiti for types A, B, and C, were 2,086, 476, and 592 respectively. In relative terms, 19.7\% of the images contain at least an occurrence of graffiti, with an approximate  66\%-15\%-19\% mixture of the types A, B, and C, respectively. That means that in the analyzed region, type A appears in two of three affected images picked at random, on the average. The dominance of this particular type on the ratios may be associated to that type being more easily produced when compared to the other types (please refer to Figure~\ref{fig:graffititypes}). The affected images may present more than one type of graffiti as can be seen in Figure~\ref{fig:coocurrence}. Despite $88\%$ of the images having occurrences of type A only, just $35\%$ of type B occur in isolation.

\begin{figure*}
    \begin{subfigure}[b]{0.32\textwidth}
	    \centering
	    \includegraphics[width=\textwidth]{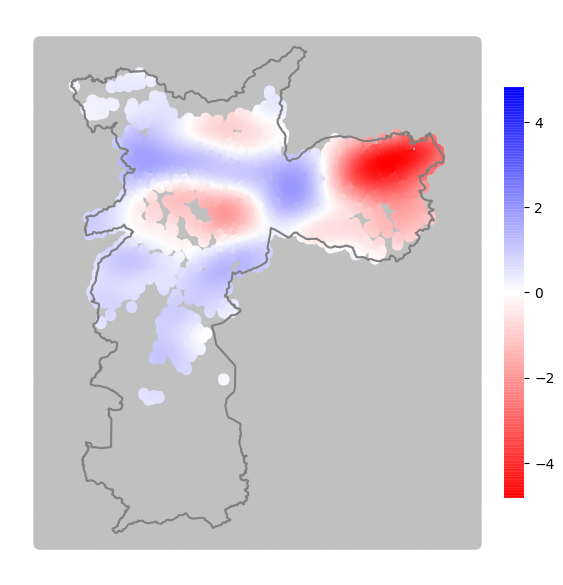}
	\caption{Type A}
    \end{subfigure}
    \begin{subfigure}[b]{0.32\textwidth}
	    \centering
	    \includegraphics[width=\textwidth]{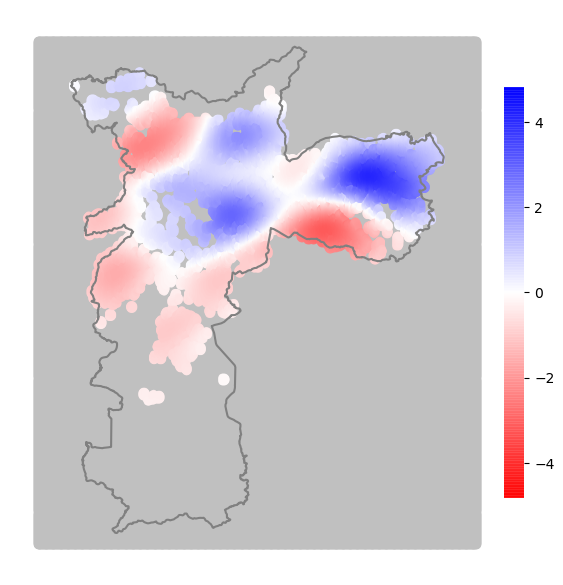}
	\caption{Type B}
    \end{subfigure}
    \begin{subfigure}[b]{0.32\textwidth}
	    \centering
	    \includegraphics[width=\textwidth]{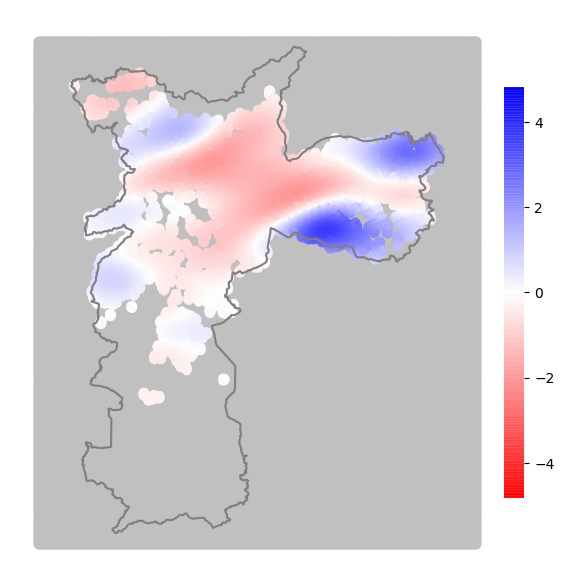}
	\caption{Type C}
    \end{subfigure}
	\caption{Graffiti occurrences distribution compared to the mean. The heatmaps show the difference between the distribution of each type and the mean distribution. The region in grey correspond to sparsely sampled locations (below one occurrence of graffiti per 500m).}
  \label{fig:maptypes}
\end{figure*}

A kernel density estimation with a gaussian kernel was used to estimate the spatial distribution of the occurrences per type. Given our interest in getting the difference among these patterns, in Figure~\ref{fig:maptypes} we depict the difference of the distribution of each type with respect to the mean distribution. Different patterns are observable for each type. It can be noticed that the northeast region of the map presents a low concentration of type A (red region in (a)) and a high concentration of type B (blue region in (b)). Also, two focuses of high concentration of type C can be observed in the eastern region (c). To discriminate between zero-valued points and sparse regions, regions with densities below one occurrence of graffiti per 500m (0.31mi) were filtered out from the analysis and in Figure~\ref{fig:maptypes}, they are shown in grey. The KL divergence between each distribution and the mean distribution was calculated as a means to quantify the discriminability
of these patterns and $0.0137$, $0.0141$, and $0.009$ were obtained for the distributions of type A, B, and C, respectively. It shows that the spatial distribution of type A differs the most from the average case.

\subsection{Analyzing the network structure}
\begin{figure}[ht]
	\centering
	\includegraphics[width=0.65\textwidth]{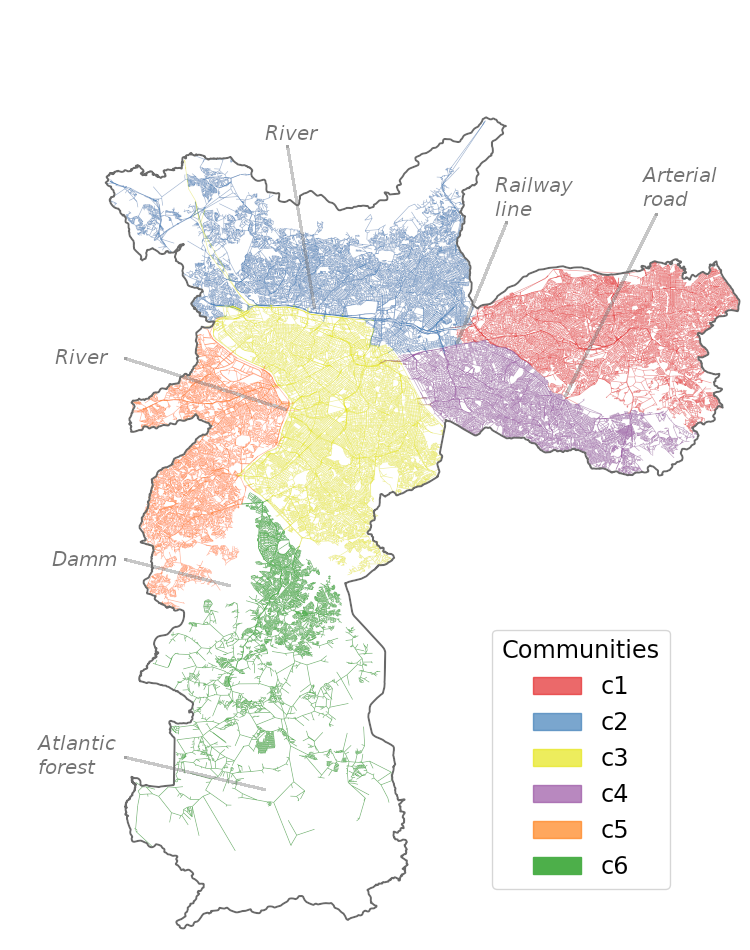}
	\caption{Network communities obtained using Infomap~\cite{rosvall2008maps}. The annotations show the consistency of the results obtained with the populated portions of the environment.}
	\label{fig:mapcomm}
\end{figure}

The application of the community methodology described in~\cite{rosvall2008maps} resulted in the six network communities (c1--c6) shown in Fig.~\ref{fig:mapcomm}. The obtained communities are congruent with the identified major separators (e.g.~avenues, rivers). The figure shows the obtained communities, with urban features annotated, including the meaning of the borders obtained and particular features of each community. As can be seen, the regions of separation mainly correspond to arterial roads and rivers that cross the city, which are natural separators of the topology of the city.

\begin{figure}
    \centering
	\includegraphics[width=.65\textwidth]{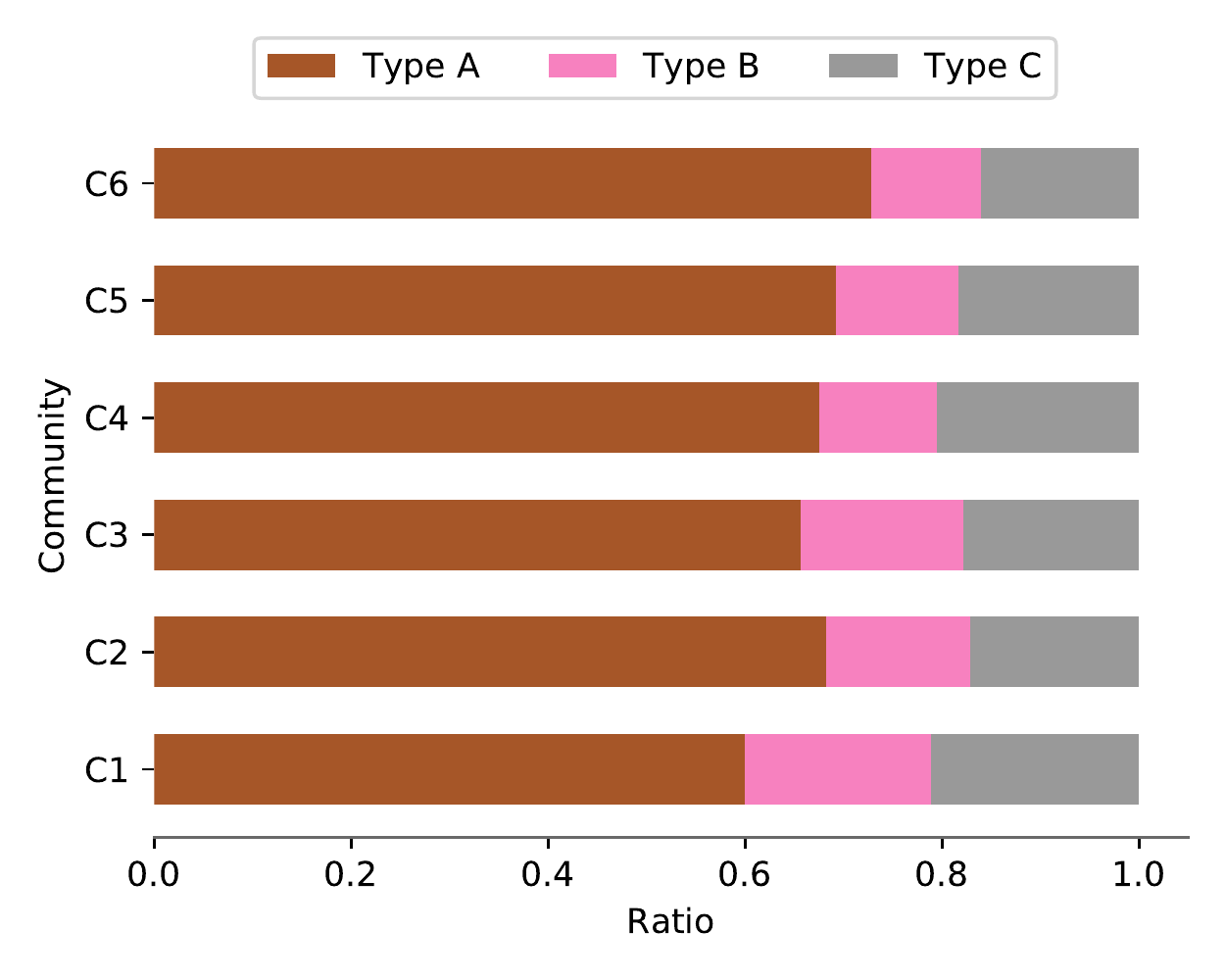}
    \caption{Ratio of types of graffiti occurrences for the obtained communities. Each bar corresponds to a community and the partitions along each bar are scaled by the ratios of each type over the number of occurrences inside that community.}
    \label{fig:typepercomm}
\end{figure}

To analyze the relationship between this structure and the graffiti distribution, the overall count of graffiti occurrences in each community was calculated. The absolute counts of all occurrences of graffiti are shown in Table~\ref{tab:percommunity}.  It is expected that the occurrences of graffiti take place predominantly in urban regions, and in fact, the table shows that the lowest concentration happens at the least urbanized region. Further analysis of the predominance of one type over another in each community can be performed considering the relative occurrence depicted in Figure~\ref{fig:typepercomm}. This chart shows a pattern common to all communities, with most of the occurrences of type A, followed by type C and B. Despite the deviations found, the results indicate that there are no significant variations on the proportions of each type of graffiti associated to the communities.

\begin{table}[ht]
    \centering
	\begin{tabular}[b]{rrrrrr}
		\toprule
		c1 & c2 & c3 & c4 & c5 &c6\\
		679 & 752 & 619 & 462 & 398 & 159\\
		\bottomrule
    \end{tabular}
    \caption{Count of graffiti per community.}
    \label{tab:percommunity}
\end{table}

\subsection{Analyzing the dynamics}
\begin{figure}
	\includegraphics[width=.65\textwidth]{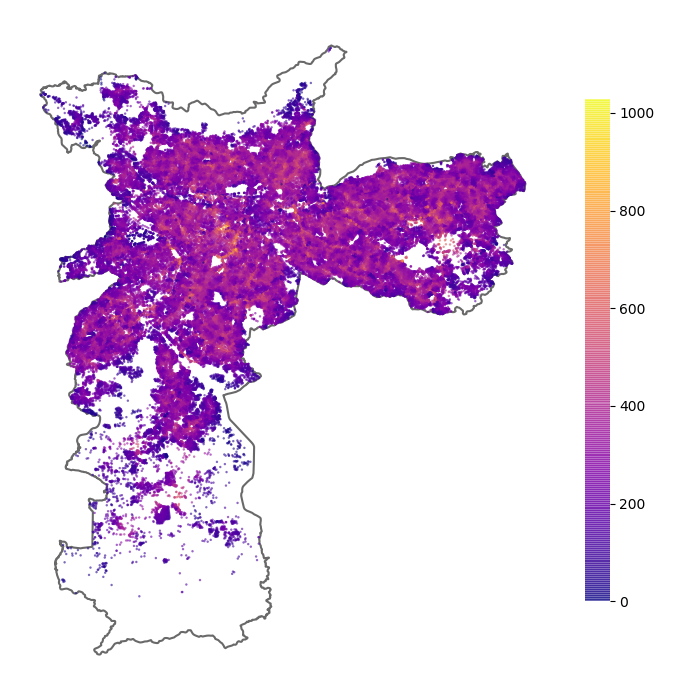}
  \caption{Accessibility of the nodes. Edges are not shown in order not to clutter the image.}
  \label{fig:accessibmap}
\end{figure}

A 20-step random-walk ($h=20$) was considered for the computation of accessibility for the entire network and the results can be seen in Figure~\ref{fig:accessibmap}. The figure shows the vertices of the network colored by the accessibility measure. It can be noticed that lower values tend to appear at the border, while the peak values occur inside the identified regions. That means that, in general, the border regions can reach less regions than the central regions
for the adopted topological scale of 20 steps.

\begin{figure}
    \centering
	\includegraphics[width=.65\textwidth]{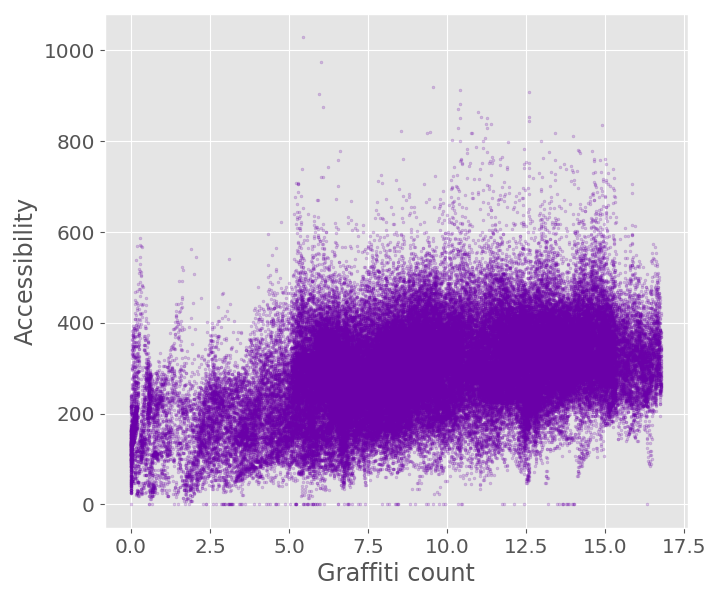}
    \caption{Relationship between the network accessibility and the overall count of graffiti occurrences. Each point represents a vertex in the network and the correspondent graffiti occurrence and the accessibility of the vertex in the network.}
    \label{fig:accessibcorr}
\end{figure}

To analyze the relationship of this vertex-based measurement and the spatial distribution of the graffiti occurrences, the graffiti count of each vertex was extracted from the previously obtained density. Vertices in sparse regions (grey regions in Figure~\ref{fig:maptypes}) were not considered. Figure~\ref{fig:accessibcorr} shows the graffiti count and the accessibility measure for each vertex of the network. The Pearson correlation coefficients between each type and the accessibility values were computed and $0.34$, $0.35$, and $0.30$ of correlation were obtained. These results indicate small positive correlation between the accessibility value and the graffiti occurrences.

In the light of these correlations, we hypothesize that the accessibility is associated to the location of graffiti in two antagonistic ways: more accessible regions are more transited and thus gets more visibility for the produced content; and at the same time, this accessible ways are more monitored, which inhibit the production of graffiti in these regions.

\section{Concluding remarks}
Graffiti is a common feature of many cities worldwide.
Some previous works have already analyzed graffiti spatial distribution, but they typically do not consider the different visual forms of graffiti. Besides, they do not take into account the city network information in the analysis, which would allow the respective topology to be tentatively associated with dynamical aspects such as mobility. 

In this work, a network science-based method for analyzing the different types of graffiti, which are then studied by considering the accessibility measurement, is reported. A case study was performed considering the city of São Paulo, Brazil and three graffiti types. The results showed that there are differences among the spatial distributions. The relative participation of each type of graffiti across the regions, though, did not change significantly. The moderate positive correlation values between the accessibility on the network and the graffiti types indicates that the spatial distributions are not strongly influenced by the process analyzed.

\section*{Acknowledgments}
The authors would like to thank FAPESP (\#2019/01077-3, \#2015/22308-2, \#2019/16223-5, \#2018/09125-4 and \#2018/10489-0), CAPES and CNPq (grant no. 307085/2018-0); NSF awards CNS-1229185, CCF-1533564, CNS-1544753, CNS-1730396, CNS-1828576; the Moore-Sloan Data Science Environment at NYU, DARPA D3M and C2SMART. Any opinions, findings, conclusions and recommendations expressed in this material are those of the authors.

\bibliographystyle{plain}
\bibliography{references}
\end{document}